\documentstyle{article}
\begin {document}
\input JAmatport
\newtheorem{propo}{Proposition}
\begin{titlepage}
\vspace*{30.0mm}
\begin{center}
{\bf \LARGE   Conformal Foliations and Constraint Quantization}\\
\medskip
\medskip
\medskip
{\large J.N. Tavares}\\
\medskip
\medskip
\medskip
 Dep. Matem\'atica Pura, Faculdade de Ciencias, U. Porto, 4000
Porto\\
\medskip

\medskip
\medskip
\medskip
{\Large Abstract}\\
\medskip
\medskip
We show that the {\it Classical Constraint Algebra} of a {\it
Par\-am\-etri\-zed
Relativistic Gauge System} induces a natural structure of {\it Conformal
Foliation} on a {\it Transversal Gauge}. Using the theory of Conformal
Foliations, we provide a natural {\it Factor Ordering} for the Quantum
Operators
associated to the Canonical Quantization of such Gauge System.\\
\medskip
\medskip
\medskip
{\Large Keywords}\\
\medskip
\medskip

Parametrized Relativistic Gauge System, Conformal Foliations, Canonical
Quantization.

\medskip
\medskip
\medskip

\end{center}
\end{titlepage}

\medskip
\medskip
\medskip

\section{Introduction}

The purpose of this paper is to show that the theory of ``{\it
Conformal Foliations}", as developed a few years ago by Montesinos,
Vaisman, and others, helps the understanding of ``{\it Constraint
(Dirac) Quantization} of  {\it Parametrized Relativistic Gauge
Systems}", by providing a natural solution for the so called ``{\it
Factor Ordering Problem}" in the ``{\it Quantum Operators}"
corresponding to the ``{\it Classical Constraints}" $H_\alpha=0=H$ (see
discussion in [8] and [5]).

We follow H\'ajic\~ek/Kucha\v r's philosophy, developed in a series of papers
(see [7],[8],[5],[6]), but here we adopt modern differential geometric
methods (in the spirit of [12] or [13]) to construct the natural quantum
operators and deduce the corresponding ``{\it Commutator Algebra}".
We hope that the use of the theory of conformal foliations helps to clarify
the essential unique geometric character of H\'ajic\~ek/Kucha\v r's
quantization method, as well as the naturality in the choice of the quantum
operators.

The paper is organized as follows. In section 2, we redefine the model,
introduced in [5], of
``{\it Parametrized Relativistic Gauge System}"  (where it will
be called only a ``{\it Gauge System}", for simplicity), using sympletic
geometry (see [1], [3] or [4]). Section 3 discusses the ``{\it Transversal
Geometry}" of the ``{\it Gauge Foliation} ${\cal F}$", in terms of a choice of
a transversal distribution $\cal T$, and shows that the mathematical
structure deduced from the classical ``{\it Constraint Algebra}"
(\ref{9})-(\ref{10}), is that of a conformal foliation with ``{\it
Complementary Form} $\lambda = -\Omega$" (see definition 4.1). Section 4
reviews the geometry of conformal foliations, following closelly [10] and
[14]. We introduce a ``{\it Conformal Curvature Tensor}" ${\bf C}$ (see
(\ref{52})) and the unique g-Riemannian connection $\nabla$, on $({\cal
T},g)$ such that ${\bf C}$ is conformally invariant (see Theorem 4.4). In
particular, this seems to be the connection introduced in [6], using a
substantial different formalism. Then, we define a
 ``{\it Scalar Curvature}" ${\bf S}$ (see definition 4.11), a ``{\it
Transversal Laplacian}" (see definition 4.13), constructed from the above
mentioned connection, and deduce some useful identities (see Lemmas 4.8, 4.10
and 4.12). Finally, in section 5, we implement the ``{\it Constraint (Dirac)
Quantization Program}", defining the quantum operators and computing the
``{\it Quantum Commutator Algebra}" (see Theorems 5.1 and 5.4). The more
relevant formulas are deduced in an appendix, at the end of the paper.

\section{Sympletic Geometry of Gauge Systems}

The situation we have in mind is the following. Let $M$
be a $(N+1)$-dimensional smooth manifold, $T^{*}M$
its cotangent bundle, together with its canonical
sympletic form $\omega = d\theta$, and
$(C^{\infty}(T^*M),\{,\})$ the Poisson algebra of
$C^\infty$ functions on $T^{*}M$.

Assume that we have a $v$-dim. integrable distribution
$\cal D$ (a subundle of $TM$) and let $\cal F$ be the
corresponding foliation on $M$. The leaves of $\cal F$
describe ``{\it Gauge Equivalent}" or physically indistinguible
configurations on $M$. We call the pair $(M,{\cal F})$ a
``{\it Gauge System}". When $\cal F$ is simple, then $ m =
M/{\cal F}$ is called the physical space and $T^{*}m$ the
physical phase space of the gauge system.

We consider now a special subalgebra of $(C^{\infty}(T^*M),\{,\})$,
consisting of functions ``{\it at most Linear in Momenta}", linearly
generated by the following two types of functions:

\medskip
   (i)... {\it Configuration functions} - functions  $F\in
C^\infty(M)$, identified with their pull-backs to $T^{*}M$.

\medskip
   (ii)... {\it Momentum functions} - to each vector field
$X\in{\cal X}(M)$ we associate the corresponding ``{\it Momentum Function}"
$J_X$ defined by:

\begin{equation}
J_X(\alpha_Q)=\alpha_Q(X_Q)  \quad \forall \alpha_Q\in T^{*}M \label{1}
\end{equation}

\medskip

  The following Poisson brackets characterize the kinematics of
the system:

 \begin{propo}
 ([1])

\medskip

(i)...  $\{F_1,F_2\} = 0$

(ii)...  $\{F,J_X\} = X^.F$

(iii)...  $\{J_X,J_Y\}= -J_{[X,Y]}$

\medskip

$\forall F_1,F_2 \in C^\infty(M) \ \ \ \forall X,Y \in
{\cal X}(M)$.

\end{propo}

  In particular, we consider momentum functions of the form $J_X$
with $X\in\Gamma\cal D$ (i.e., $X$ is a vector field on
$M$, tangent to $ \cal F$) which we call ``{\it Constraint Functions}".
They form a subalgebra of $(C^{\infty}(T^*M),\{,\})$ called the
(linear) ``{\it Gauge Algebra}" $\cal G$ of the system. Using this, we
define now the ``{\it Constraint Set}" ${\cal C}\subset T^{*}M$ by:

\begin{equation}
{\cal C} = \{\alpha_Q\in T^{*}M : J_X(\alpha_Q) = 0, \forall X \in
\Gamma \cal D\} \label{2}
\end{equation}

Note that $\cal C$ is the subundle ${\cal D}^{o}$ of $T^{*}M$,
consisting of the covectors that vanish on $\cal D$ and, therefore,
that $\cal C$ is a coisotropic submanifold in $T^{*}M$ (a first
class constraint set, in Dirac's terminology). When we consider
the closed 2-form $i^{*}\omega$ (where $i:{\cal C}\to T^{*}M$ is
inclusion), we see that it is degenerate and its characteristic
distribution ${\cal K } = Ker (i^{*}\omega)$ is integrable and
locally generated by the Hamiltonian vector fields associated to
the constraint functions. The quotient (if it exists) of $\cal C$
by the characteristic foliation determined by $\cal K$, ${\cal S}
= {\cal C}/{\cal K}$ has a unique sympletic form  $\omega_{\cal
S}$ such that $\pi^{*}\omega_{\cal S} = i^{*}\omega$ (see [1],[3] or [4]):

$$\begin{array}{rcl}
{\cal C}& \stackrel{i} \longrightarrow &T^*M\\
                    \mapdown{\pi}& & \\
                    {\cal S} = {\cal C}/{\cal K}& &
\end{array}$$
Also if $ F_1,F_2 \in C^{\infty}(T^{*}M)$ are two functions
constant on the leaves of the characteristic foliation, i.e., such
that $i^{*}F_k = {\pi}^{*}f_k$ (k=1,2), for uniquely determined
functions $f_1,f_2 {\in} C^\infty(\cal S)$, then:

\begin{equation}
i^{*}\{F_1,F_2\} = {\pi}^{*}\{f_1,f_2\}_{\cal S} \label{3}
\end{equation}
Under certain regularity assumptions, we know that $({\cal
S},\omega_{\cal S})$ is symplectomorphic to $(T^{*}
m,{\omega}_ m=d{\theta}_m)$ and so, it is the natural candidate for
the physical (or reduced) phase space of the physical system (see [3]).

Now we assume that the intrinsic dynamics of our gauge system is generated
by a quadratic function in momenta, of the form:

\begin{equation}
H = 1/2 \, G + J_U +V  \label{4}
\end{equation}
where G is a contravariant ``metric" (eventually degenerate) on
$M$, viewed as a function on $T^*M$, $U\in{\cal X}(M)$ is a vector field in M
and
$V\in{\cal C}^{\infty}(M)$. We have the following Poisson brackets,
involving the new kind of homogeneous quadratic function G:

\begin{propo}
([1])

\medskip

(i)...  $\{G,F\} = -2 J_{grad_G F}$, where $grad_G F$ is the vector
field $G(dF,.) \in {\cal X}(M)$

(ii)...  $\{G,J_X\} ={ L_X}G $, the Lie derivative of $G$ in the
direction of $X \in {\cal X}(M)$
\end{propo}

$\forall F \in C^\infty(M) \ \ \ \ \ \forall X \in {\cal X}(M)$.
\medskip

Now we consider the equations of motion, which follow from the canonical action
principle, for a parametrized gauge system:

\begin{equation}
S(Q^{A},P_{A};N,N^{\alpha}) = {\int} dt
(P_{A}{\dot Q}^{A}-N.H-N^{\alpha}.H_{\alpha})\ \longrightarrow\  stat.
\label{5}
\end{equation}
where $(Q^{A},P_{A})$ are local canonical coordinates in $T^{*}M$; $N$,
$N^{\alpha}$ Lagrange multipliers and $H_{\alpha} =
J_{X_{\alpha}}$ are momentum functions associated to a local basis
$X_{\alpha}$ (${\alpha} = 1,\ldots,v$) of the ``{\it Gauge Distribution}"
$\cal D$ on M.

Following the physical terminology, we call the $H_\alpha$ the
``{\it Supermomentum Functions}" and $H$ the ``{\it Superhamiltonian}" of the
(parametrized) Gauge system (see [5] for discussion and examples).

Variation with respect to $Q^A$ and $P_A$ yields the Hamiltonian
equations, while variation with respect to the ``{\it Lapse Function}"
$N$ and the ``{\it Shift Vector Field}" $N^\alpha$, leads to the
constraints:

\begin{equation}
H = 0 = H_\alpha \label{6}
\end{equation}

So, we recover the ``{\it Kinematical Constraint Set}" $\cal C$, defined
in (2), together with a new ``{\it Dynamical Constraint}" $H = 0$, that
reveals the fact that the system is invariant under external ``time"
reparametrizations.

Assume that ${\cal C}_o = H^{-1}(0)$ is a
submanifold of $T^{*}M$. Since codim ${\cal C}_o = 1$, ${\cal C}_o$  is
coisotropic and is foliated by the trajectories of $X_H$ (the Hamiltonian
vector
field associated to $H$) that represent the evolution of the
system. For consistency, we assume that the full set of constraints
(\ref{6}) is preserved by the dynamics, which means that ${{\cal C}_o} \cap
{\cal C}$ is coisotropic, i.e., we have:

\begin {equation}
\{H,H_\alpha\} = C_\alpha H + C_\alpha^\beta H_\beta \label{7}
\end {equation}
where $H_\alpha = J_{X_\alpha}$ are supermomenta associated to a local basis
$X_\alpha$ of $\cal D$, $C_\alpha \in C^\infty(M)$ and $C_\alpha^\beta  \in
C^\infty(T^{*}(M))$ is at most linear in momenta. If the
local basis $X_\alpha$ for $\cal D$ verifies:

\begin{equation}
[X_\alpha,X_\beta] = - \Gamma_{\alpha\beta}^\gamma X_\gamma  \label{8}
\end{equation}
with $\Gamma_{\alpha\beta}^\gamma \in
C^\infty(M)$, then the ``{\it Algebra of Constraints}" $H=0=H_\alpha$ has
the following (open) structure:

\begin{equation}
\{H_\alpha,H_\beta\} =
\Gamma_{\alpha\beta}^\gamma H_\gamma \label{9}
\end{equation}
\begin{equation}
\{H,H_\alpha\} = C_\alpha H + C_\alpha^\beta H_\beta    \label{10}
\end{equation}
$H$ is given by (\ref{4}) and, working  equality (\ref{10}), using
propositions 1 and 2, we conclude the following identities for the Lie
derivatives in ``{\it Gauge Directions}":

\begin{equation}
L_\alpha G = C_\alpha G \hbox{    (mod {\bf I})}  \label{11}
\end{equation}
\begin{equation}
L_{\alpha} U = C_\alpha U  \hbox{   (mod {\bf I})}  \label{12}
\end{equation}
\begin{equation}
L_\alpha V = C_\alpha V  \label{13}
\end{equation}
where {\bf I} is the ideal of the contravariant tensor algebra $\bigotimes
TM$, generated by $\Gamma\cal D$ and $L_\alpha$ is Lie derivative in the
direction of $X_\alpha \in \Gamma\cal D$. So, this Lie derivative rescales
the fields by a common factor and adds certain elements from {\bf I}.

All the conclusions about the physical content of the gauge system described
 by the above model,   must  be deduced only from intrinsic data, namely, from
the ``{\it Kinematical Constraint} ${\cal C}$", determined by the foliation
${\cal F}$, and the ``{\it Dynamical Constraint} ${\cal C}_o$", given by
$H=0$. So, they must be invariant under transformations that preserve these
data. We impose also that these transformations preserve the polynomial
character (in momenta) of the constraints (this will be needed for the
quantization program, as we shall see later, and is related to Van Hove's
theorem (see [1], section 5.4)). So, they must be a combination of the
following three types of transformations:

\medskip
($\bf A$). {\it Change in the local basis for $\cal D$}:
\begin{equation}
X_\alpha \to \overline{X}_{\alpha}=\Lambda_{\alpha}^\beta X_\beta  \label{a1}
\end{equation}
 where $\Lambda_{\alpha}^\beta\in C^\infty(M)$, $det \,
\Lambda_{\alpha}^\beta\not= 0$, with the corresponding change in
supermomenta:
\begin{equation}
H_{\alpha} \to \overline H_{\alpha}=\Lambda_{\alpha}^\beta
H_\beta   \label{a2}
\end {equation}

\medskip
($\bf B$). {\it  Gauging the superhamiltonian}:
\begin{equation}
H \to \overline{H}={H}+\Lambda^\alpha {H}_\alpha       \label{a3}
\end {equation}
 with $\Lambda^\alpha\in C^\infty(T^*(M))$,  at most linear in momenta.

\medskip
($\bf C$). {\it  Scalling the superhamiltonian}:
\begin {equation}
H \to\overline{H}=e^\Omega {H}  \label{a4}
\end {equation}
 onde $\Omega \in C^\infty(M)$. To preserve the signature of the metric we
only allow positive scallings.

Note that the above transformations doesn't leave invariant the
superhamiltonian $H$, given by (\ref{4}). In fact, after an easy computation,
we deduce that, if:
$${H}\to \overline{H}=\frac{1}{2} \overline{G}
+J_{\overline {U}}+\overline{V}$$
then:
\begin {eqnarray}
\overline{G} &=& e^\Omega(Q) G \ \ (mod \,{\bf I}) \label{a10}\\
\overline{U} &=& e^\Omega(Q) U \ \ (mod \,{\bf I}) \label{a11}\\
\overline{F} &=& e^\Omega(Q) V  \label{a12}
\end {eqnarray}

  Let us say that two superhamiltonians ${H}=\frac {1}{2} G +J_U +V$ and
 $\overline {H}=\frac {1}{2} \overline G +J_{\overline U} + \overline V$,
 are {\it conformally equivalent} (mod  {\bf I}), if they are related through
 (\ref{a10}) to (\ref{a12}).

Then we conclude that  the above transformations ({\bf
A},{\bf B} and {\bf C}), leave invariant the conformally equivalence
class (mod {\bf I})  of the superhamiltonian.

\medskip

\section{Transversal Geometry of $\cal F$}

The transversal geometry of $\cal F$ is infinitesimally modeled by the
normal bundle $TM/\cal D$. We make a choice of a transversal
subundle $\cal T$ to $\cal D$:

\begin{equation}
T_{Q}M = {\cal D}_Q \oplus {\cal T}_Q  \label{15}
\end{equation}
which identifies $TM/{\cal D} \cong {\cal T}$. Vectors on $\cal D$ are
called ``{\it Longitudinal}", and vectors on $\cal T$ are called
``{\it Transversal}". Associated to (\ref{15}), we have two projectors:

\begin{equation}
Id = P_{\parallel} \oplus P_{\perp} \label{16}
\end{equation}
a decomposition:
\begin{equation}
{T_Q}^*M = {{\cal D}_Q}^* \oplus {{\cal T}_Q}^* \label{17}
\end{equation}
with ${{\cal T}^{*}_Q} \cong {\cal D}^{o}_Q$, and the two corresponding
projectors:
\begin{equation}
Id^* = {P_\parallel}^* \oplus {P_\perp}^*  \label{18}
\end{equation}

We can form various tensor products of these projectors, using them to
project the various tensors into longitudinal and transversal
parts. In particular, the choice of the ``{\it Transversal Gauge}" $\cal T$
allows us to define the ``{\it Transversal Superhamiltonian}" $H_{\perp}$
by:

\begin{equation}
H_{\perp} = 1/2 \, G_{\perp} + J_{U_\perp} + V  \label{19}
\end{equation}
where:

\begin{equation}
G_{\perp} = (P_{\perp} \otimes P_{\perp})(G) \label{20}
\end{equation}
and

\begin{equation}
U_{\perp} = P_{\perp} U \label{21}
\end{equation}
are the ``{\it Transversal Contravariant Metric}"  and the ``{\it
Transversal Vector Potencial}", respectivelly (associated to $G$ and $\cal
T$).

So, the transversal gauge $\cal T$ fixes a representative of the
equivalence class of superhamiltonians, connected by the ``{\it Gauging}"
({\bf B}), since the difference between two transversal projections
of the same vector, belongs to $\cal D$.

Let us compute the constraint algebra in the transversal gauge $\cal T$.
For this, we compute first $L_\alpha G_\perp$ and $L_\alpha U_\perp$,
using the fact that $L_\alpha$ is a tensor derivation that commutes with
contractions:

\begin{equation}
{L_\alpha}({\cal C}({\cal P}_\perp \otimes t))  = {\cal
C}({L_\alpha}{\cal P}_\perp \otimes t) + {\cal C}({{\cal
P}_\perp}\otimes{L_\alpha}t) \label{22}
\end{equation}
where $\cal C$ is a contraction, ${\cal P}_\perp$ a transversal projector
(a tensor product of some ${P_\perp}$ and ${P_\perp}^*$) and t some tensor
field on M.

Define for each  $\alpha,\beta\in\{1,...,v\}$ a transversal 1-form (i.e., a
form which anihilates longitudinal vectors) $\omega_\alpha^\beta$ by:

\begin{equation}
\omega_\alpha^\beta = (L_\alpha P_\perp^*)(\theta^\beta)  \label{23}
\end{equation}
where $\theta^\beta\in\Gamma{\cal D}^*$ is the dual basis to
$X_\beta\in\Gamma{\cal D}$. A computation, using local frame fields for
$\cal D$, $\cal T$, their duals and also formula (\ref{22}) (see
Appendix), shows that:

\begin{eqnarray}
L_\alpha U_\perp &=& C_\alpha U_\perp + {\omega_\alpha^\beta}(U_\perp)X_\beta
\nonumber \\
&=&C_\alpha U_\perp \ \ (modJ\,{\bf I}) \label{24}
\end{eqnarray}

\begin{eqnarray}
L_\alpha G_\perp &=& C_\alpha G_\perp + {V_\alpha^\beta}\vee X_\beta
\nonumber \\
&=&  C_\alpha G_\perp \ \ (modJ\, {\bf I})   \label{25}
\end {eqnarray}
where ${V_\alpha^\beta}=G({\omega_\alpha^\beta},.)$ is the transversal
vector field ``{\it metric-equivalent}" to ${\omega_\alpha^\beta}$, and
$\bigvee$ denotes symmetric tensor product.

Finally, using (\ref{24}) and (\ref{25}) together with proposition 1 and 2, we
deduce that:

\begin{equation}
\{H_\perp,H_\alpha\} = C_\alpha H_\perp + F_\alpha^\beta H_\beta \label{27}
\end{equation}
where $F_\alpha^\beta \in C^\infty(T^*M)$ is given by:
\begin{equation}
F_\alpha^\beta = J_{V_\alpha^\beta} - \omega_\alpha^\beta (U_\perp) \label{28}
\end{equation}

Now we arrive at a crucial point - the definition of a 1-form
$\Theta$, on M, attached to the geometry of the gauge system as specified
by the constraint algebra (\ref{9}) and (\ref{10}). So, consider the
longitudinal 1-form:
\begin{equation}
\Theta = C_\alpha \theta^\alpha  \label{29}
\end{equation}
where $\theta^\alpha \in \Gamma{\cal D}^*$ are dual 1-forms to the
$X_\alpha$. It's easy to see that $\Theta$ is a globally well-defined
longitudinal 1-form on $M$. We want to compute its ``{\it Foliated
Derivative}" $d_{\cal F}\Theta$ (i.e., the derivative along longitudinal
directions, see [16]). For this, we first note that Jacobi Identity:
\begin{equation}
cycl sum \{H,\{H_\alpha,H_\beta\}\} = 0 \label{30}
\end{equation}
implies the following identity:
\begin{equation}
\{C_\alpha,H_\beta\} - \{C_\beta,H_\alpha\} = \Gamma_{\alpha\beta}^\gamma
C_\gamma  \label{31}
\end{equation}

 So, recalling that $C_\alpha, \Gamma_{\alpha\beta}^\gamma  \in
C^\infty(M)$, we deduce that:
\begin{equation}
L_\beta C_\alpha - L_\alpha C_\beta = \Gamma_{\alpha\beta}^\gamma
C_\gamma  \label{32}
\end{equation}
and finally:
\begin{eqnarray*}
d_{\cal F}\Theta(X_\alpha,X_\beta)&=& d\Theta(X_\alpha,X_\beta)\\
&=&
X_\alpha.\Theta(X_\beta) - X_\beta.\Theta(X_\alpha) -
\Theta([X_\alpha,X_\beta])\\
& =& L_\alpha C_\beta - L_\beta C_\alpha -
\Theta(- \Gamma_{\alpha\beta}^\gamma X_\gamma)\\
&=& L_\alpha C_\beta - L_\beta
 C_\alpha + \Gamma_{\alpha\beta}^\gamma C_\gamma\\
& =& 0
\end{eqnarray*}
i.e., $\Theta$ is $d_{\cal F}$-closed. Recall that a kind of
``{\it Poincar\'e Lemma}" (see [15]) applies to this case - locally, there
exists a function $\Omega$ such that $\Theta = d_{\cal F}\Omega$, i.e.:
\begin{equation}
\Theta(X) = {L_X}\Omega,\  \forall X\in \Gamma{\cal D}  \label{33}
\end{equation}

In particular, for $X=X_\alpha$, we have:

\begin{equation}
C_\alpha = {L_\alpha}\Omega  \label{34}
\end{equation}

Consider now the ``{\it Transversal Contravariant Metric}" $\tilde
g=G|_{{\cal D}^o} = {G_\perp}|_{{\cal D}^o}$ (see Appendix), that we assume
to be nondegenerate. Let $g$ be the associated transversal covariant
metric on $\cal T \cong {\cal T}^{**}$ $\cong$ ${{\cal D}^o}^*$. A
computation made in the Appendix (see also the note following definition 4.1),
shows that:

\begin{equation}
L_\alpha g = - \Theta(X_\alpha) g  \label{35}
\end{equation}

\medskip

Find a function $\Omega$ that locally verifies (\ref{34}), and define the
rescalled metric $\overline {g}$ by:
\begin{equation}
\overline {g} = e^{\Omega}g    \label{36}
\end{equation}

Then we have:
\begin{eqnarray*}
{L_\alpha}\overline{g} = L_\alpha (e^\Omega g) &=&
(L_\alpha \Omega)\overline{g} + e^\Omega L_\alpha g \\&=& C_\alpha
\overline{g} - C_\alpha \overline{g}\\&=& 0
\end{eqnarray*}
and we see that the rescaled metric is foliated (i.e., constant along
the leaves of $\cal F$), or, put another way, $g$ is locally conformal to
a foliated metric.

Acording to Montesinos ( see [10],[11] and section 4) we say that $\cal F$ is a
``{\it Conformal Foliation}" and that:

\begin{equation}
\lambda = - \Theta  \label{37}
\end{equation}
is the corresponding ``{\it Complementary Form}".

\medskip

One more point, before closing this section.

Recall that a choice of a transversal subundle $\cal T$, fixes a
representative $H_\perp$ in the equivalence class of superhamiltonians
connected by the ``{\it Gauging Transformations}" ({\bf B}).
However, we are still free to ``{\it rescale}" the superhamiltonian acording
to ({\bf C}). When we do this, and compute the new structure
functions in (\ref{10}), we see that:
\begin{equation}
C_\alpha \to \overline{C_\alpha} = C_\alpha + L_\alpha \Omega  \label{38}
\end{equation}
and
\begin{equation}
C_\alpha^\beta \to \overline{C_\alpha^\beta} = e^\Omega C_\alpha^\beta
\label{39}
\end{equation}

\medskip

Hence we conclude two things: first, the $d_{\cal F}$-cohomology class
$[\lambda]$ remains unchanged. In fact:
$$\overline{\Theta} - \Theta =
(\overline{C_\alpha} - C_\alpha)\theta^\alpha = (L_\alpha
\Omega)\theta^\alpha = d_{\cal F}\Omega$$

  Secondly, as $\lambda=-\Theta$ is $d_{\cal F}$-closed, locally we can
choose a function $\Omega$ such that $d_{\cal F}\Omega=\lambda$,
which implies in particular that $d_{\cal F}\Omega(X_\alpha)=L_\alpha
\Omega=\lambda(X_\alpha)=-C_\alpha$, and so, by (\ref{38}),
$\overline{C_\alpha} = 0$, for the corresponding rescalled superhamiltonian
$\overline{H}=e^\Omega H$.

If we can find globally such a $\Omega$, i.e., if
$\lambda$ is $d_{\cal F}$- exact, then:
 \begin{equation}
L_\alpha \overline{G} = 0 =
L_\alpha\overline{U} = 0 = L_\alpha \overline{V} (mod\, {\bf I}) \label{a20}
\end{equation}
and so the
fields $\overline{G},\overline{U},\overline{V}$ are projectable in the (leaf)
physical space (when it exists). However, note that $\Omega$ is defined up to
a function $\omega$ such that $d_{\cal F} \omega=0$, i.e., up to a {\it basic
function} (constant on the leaves of the foliation $\cal F$).

 So, when the foliation is simple, we will have a physical superhamiltonian
$h$, defined up to a multiplication by a function $e^w$, with $w \in
C^\infty(M)$ a basic function. In other words, the gauge system only
determines the conformal geometry in the physical configuration space.
Moreover, in this case, if $\pi: {M} \to m={M}/{\cal F}$, denotes the canonical
projection, then, as ${\pi}_{*}=d\pi$ anihilates {\bf I},  we obtain a
conformally class of contravariant tensors on $m$, ($e^w.g, e^w.P_u, e^w.v$),
$w\in C^\infty (m) $, and also conformally class of ``physical
superhamiltonians":
\begin{equation}
\{h\} = \{e^w(g+P_u+v) : w\in\ C^\infty (m) \} \label{a30}
\end{equation}

\section{Geometry of Conformal Foliations}

In the last section, we have seen that the gauge system determines the
structure of conformal foliation on $\cal F$. To be more specific, we
adopt the following definition from [10-11]:

\subsection{Definition}

 {\it Let $M$ be a manifold, $\cal D$ an integrable distribution, $\cal T$
a transversal subundle to $\cal D$ and $g$ a covariant metric on $\cal T$.
We say that ($M$,$\cal D$,$\cal T$,$g$) is a Conformal Foliation, if
there exists a longitudinal 1-form  $\lambda$ such that}:

\begin{equation}
L_X g = \lambda (X) g    \quad  \forall X \in \Gamma {\cal D} \label{40}
\end{equation}
$\lambda$ {\it is called the ``Complementary Form"  of} $\cal F$.

\medskip

{\bf Note}.

\medskip

{\small In equation (\ref{40}), $L_X g$, $X \in \Gamma {\cal D}$, is
defined as in equations (\ref{88}-\ref{89}) of the appendix.}

\medskip

Now we collect some facts about conformal foliations (see [10-11] and
[14]).

Let $\nabla$ be a linear connection on the vector bundle $\cal T$. We
define as usual its curvature and torsion, respectivelly, by:

\begin{equation}
{\bf R}(U,V)Q = ([\nabla_U,\nabla_V] - \nabla_{[U,V]})Q \label{41}
\end{equation}

\begin{equation}
{\bf T}(U,V) = \nabla_U V_\perp - \nabla_V U_\perp - [U,V]_\perp \label{42}
\end{equation}
where $U,V \in{\cal X}(M)$ and $Q\in \Gamma\cal T$.

Define ${\cal D}^{p,q}$ as the space of (p,q)-double forms on $M$, i.e.,
the space of p-forms on $M$ with values on transversal q-forms (which
anihilate vectors on $\cal D$):
\begin{equation}
{\cal D}^{p,q} = {\bigwedge ^p T^{*}M}\otimes{\bigwedge^q {\cal T}^*}
\label{43}
\end{equation}

With ${\bf R}$ and ${\bf T}$ we associate two double forms ${\bf K}\in{\cal
D}^{2,2}$ and ${\bf N}\in{\cal D}^{2,1}$ defined, respectivelly, by:
\begin{equation}
{\bf K}(U,V;Q,S) = g({\bf R}(U,V)Q,S) \label{44}
\end{equation}
\begin{equation}
{\bf N}(U,V;Q) = g({\bf T}(U,V),Q)  \label{45}
\end{equation}
and also the ``{\it Transversal Torsion}" ${\bf N}_\perp$, of $\nabla$ by:
\begin{equation}
{\bf N}_\perp (U,V;.) = {\bf N}(U_\perp,V_\perp;.) \label{46}
\end{equation}
$\forall U,V\in {\cal X}(M), \forall Q,S \in \Gamma\cal T$.

\subsection{Definition}

{\it We say that a linear connection $\nabla$ on the Riemannian vector
bundle $({\cal T},g)$ is {g-Riemannian}, if it verifies the
following two conditions}:

\medskip

(i)... \begin{equation}
U.g(Q,S) = g(\nabla_U Q,S) + g(Q,\nabla_U S)  \label{a40}
\end{equation}
and:
(ii)... \begin{equation}
{\bf N}_\perp = 0 \ \ \ {\hbox{i.e.}} \ \ \  \nabla_{U_\perp} V_\perp -
\nabla_{V_\perp} U_\perp - {[U_\perp,V_\perp]}_\perp = 0 \label{a41}
\end{equation}
\medskip
$\forall U,V \in {\cal X}(M), \forall Q,S \in \Gamma {\cal T}$.
(Note that we are only assuming $g$ nondegenerate).

\medskip

For example, if $\tilde{G}$ is a covariant metric on $M$, such that
$\tilde{G}|{\cal T} = g$, and if $\tilde{\nabla}$ is the Levi-Civit\`a
connection
of $\tilde{G}$, then:

\begin{equation}
\nabla_U  Q = {(\tilde{\nabla}_U Q)}_\perp \label{47}
\end{equation}
is a g-Riemannian connection on $({\cal T},g)$. So g-Riemannian
connections are not unique and this difficults the definition of a natural
conformal curvature tensor for the transversal bundle $({\cal T},g)$.

However, by a careful analysis based on early work of Kulkarni,
Montesinos succeeds in isolating a class of conformal equivalent
connections, on which he defines a conformal curvature tensor.

\medskip

{\bf Note}.

\medskip

{\small

For motivation, recall the classical theory: conformal
(Weyl) curvature is a tensor field associated to a class of conformally
equivalent Riemannian connections and which is conformal invariant:

$$\begin{array}{rcl}
{g}&\longrightarrow&
\overline{g} = e^{2f}g \\ \nabla = \nabla_g&\longrightarrow&\overline
{\nabla} =   \nabla_{\overline g},\hbox{ defined  by}\\
&& \overline
{\nabla}_X Y = \nabla_X Y +(Xf)Y+(Yf)X\\
&& -g(X,Y)grad_g f  \\
{\bf K}\hbox{ (Riemann
tensor)}&\longrightarrow&\overline {\bf K} = e^{2f}({\bf K}-....)  \\ {\bf C}
&=&\overline {\bf C}\hbox{  (Weyl tensor)}
\end{array}$$}

\medskip

Here the dificulty  is in the analogue of the second line in the above
scheme - what must be $\overline {\nabla} = \nabla {\overline g}$?
Montesinos answers the following: let $\nabla$ be any g-Riemannian
connection on $({\cal T},g)$. Then there exists a unique $
\overline{g}$-Riemannian connection $\overline{\nabla}$, given by:

\begin{equation}
\overline{\nabla}_U Q = \nabla_U Q + (Q.\Omega)U_\perp + (U.\Omega)Q -
g(U;Q)Z   \label{48}
\end{equation}

It's torsion $\overline{\bf N}$ is:

\begin{equation}
\overline {\bf N} = e^{2\Omega}({\bf N} - g \wedge(d\Omega)_\parallel)
\label{49}
\end{equation}

{\bf Note}.

\medskip

{\small

The meaning of the symbols in the above equations, is the following: $g$
is interpreted as a (1,1)-double form, defined by $g(U;Q)=g(U_\perp,Q)$;
$(d\Omega)_\parallel \in  {\cal D}^{1,0}$ is defined by
$(d\Omega)_\parallel(U) = (d\Omega)(U_\parallel) = (d_{\cal F}\Omega)(U)$;
$g\wedge(d\Omega)_\parallel \in {\cal D}^{2,1}$ is defined in the usual
way by:

$g\wedge(d\Omega)_\parallel(U,V;.) = g(U,.)(d\Omega)_\parallel(V) -
g(V;.)(d\Omega)_\parallel(U)$
and finally:
$Z = {\bf grad}_\perp{\Omega} = g^{-1}(d\Omega|{\cal T},.) \in \Gamma \cal T$
is the ``{\it Transversal Gradient}" of $\Omega \in C^\infty(M)$ (see
definition 4.13).}

\subsection{Definition}
{\it We call $\overline \nabla$, given by (\ref{48}), the Connection
Conformally Associated to $\nabla$, by the conformal scaling}:

\begin{equation}
g \to \overline {g} = e^{2\Omega} g,\  \Omega \in C^\infty(M)   \label{50}
\end{equation}

\medskip

Based on this class of conformally equivalent connections [10] proceeds
in the construction of a conformal curvature tensor, in the following
steps:

\medskip

({\bf 1}). First, he proves that, given a $w\in{\cal D}^{k+1,l+1}$ with
$b=v-k-l\geq 1$,there exists only one $\delta w \in {\cal D}^{k,l}$ such
that (see Th.3.3 in [10]):

\medskip

${\bf c}(w - g\wedge \delta w) = 0$

\medskip

Here {\bf c} means contraction: for an $\alpha \in {\cal D}^{k+1,l+1}$ we
define ${\bf c}{\alpha} \in {\cal D}^{k,l}$ by ${\bf
c}\alpha(U_1,...,U_k;Q_1,...,Q_l) = \sum_a
\epsilon_a(E_a,U_1,...,U_k;E_a,Q_1,...,Q_l)$, where $E_a$ is an
orthonormal frame field for $\Gamma{\cal T}$ and $\epsilon_a =
g(E_a,E_a)$.

With this, he defines a ``{\it Conformal Operator}" ${\bf conf}:{\cal
D}^{k+1,l+1} \to {\cal D}^{k+1,l+1}$ by:

\begin{equation}
{\bf conf}w = w - g\wedge \delta w   \label{51}
\end{equation}
so that:
\medskip
{\bf c}({\bf conf}w)=0.

\medskip

({\bf 2}). In particular, for ${\bf K}\in {\cal D}^{2,2}$ given by (\ref{44}),
we define the {\it Conformal Curvature Tensor} {\bf C} by:

\begin{equation}
g({\bf C}(U,V)Q,S) =({\bf conf}{\bf K})(U,V;Q,S) \label{52}
\end{equation}

\medskip

({\bf 3}). Making a conformal change in $g$:
$g\to\overline{g} = e^{2\Omega}g,  \Omega \in C^\infty(M)$, we can prove
that:

\medskip

(i)...  $${\bf conf} = \overline{\bf conf}$$
and that:
(ii)... $${\bf conf}\overline{\bf K} = e^{2\Omega}{\bf conf}({\bf K} -
{\bf N}\wedge d\Omega|{\cal T})$$

\medskip

So, denoting by $\overline{\bf C}$ the conformal curvature given by (\ref{52})
and constructed with $\overline\nabla$, given by (\ref{48}), we see that:

\begin{eqnarray*}
e^{2\Omega}(\overline{{\bf C}}(U,V)Q,S)&=& \overline{g}(\overline{{\bf
C}}(U,V)Q,S)\\&=& \overline{{\bf conf}}\overline {\bf K}(U,V;Q,S) \ \ \ \ \
\ \ \ by (\ref{52})\\
&=& {\bf conf}\overline{\bf K}(U,V;Q,S)  \ \ \ \ \ \  {\hbox{by (i). above}}
 \\
&=& e^{2\Omega}{\bf conf}({\bf K} - {\bf N}\wedge d\Omega|{\cal
T})(.,.;.,.) \ \ \ \  {\hbox{by (ii). above}} \\
&=& e^{2\Omega}{\bf
conf}(.,.;.,.) - e^{2\Omega}({\bf N}\wedge d\Omega|{\cal T})(.,.;.,.)
 \end{eqnarray*}
and we conclude that {\bf C} is conformally invariant ({\bf C} =
$\overline{\bf C})$ iff:

\begin{equation}
{\bf conf}({\bf N}\wedge d\Omega|{\cal T}) = 0   \label{53}
\end{equation}
$\forall \Omega \in C^\infty(M)$.

Now comes the main point, namely, the existence of a conformally
invariant conformal curvature for the transversal bundle $\cal T$ of
$\cal F$, is an exclusive property of conformal foliations (see Th.4.2
in [10]). For us, the main interest is in the following:

\subsection{Theorem}

{\it Let ($M$,$\cal D$,$\cal T$,$g$) be a conformal foliation with
complementary form $\lambda$, and assume that codim ${\cal D} = n+1 \geq
3$. Then there exists a unique g-Riemannian connection on $({\cal T},g)$
such that {\bf C} is conformal invariant}.

\medskip

{\bf Sketch of proof}...

\medskip

{\small

Let $\tilde{G}$ be any metric on $M$ such that $\tilde{G}|{\cal T} = g$, and
let
$\tilde{\nabla}$ be its Levi-Civit\`a connection. Define a connection
$\nabla$ on $({\cal T},g)$ by the formula:
\begin{equation}
{\nabla_U}Q = (\tilde{\nabla}_U Q)_\perp - (\tilde{\nabla}_Q
U_\parallel)_\perp + 1/2 \, \lambda (U) Q  \label{54}
\end{equation}
$\forall U\in{\cal X}$ and $Q \in \Gamma {\cal T}$

\medskip

We can prove the following facts:

(i)... $\nabla$ is g-Riemannian

(ii)... ${\bf N} = 1/2(\lambda\wedge g)$

(iii)... ${\bf K}_\parallel = 0$
where ${\bf K}_\parallel(U,V;..) = {\bf K}(U_\parallel,V_\parallel;..)$,i.e.,
$\nabla$ is Flat in longitudinal directions.

(iv)... $\forall X \in \Gamma{\cal D}$:

\begin{equation}
\nabla_X Q = (L_X Q)_\perp + 1/2 \, \lambda(X)Q  \label{55}
\end{equation}

\medskip

Now we deduce, by computation, that (ii). above implies (\ref{53}), so that
$\nabla$ solves our problem. Conversely, (\ref{53}) implies that the torsion
$N$ must be given by (ii). above, and so $\nabla$ is unique.////}

\subsection{Definition}

{\it We say that the conformal foliation ($M$,$\cal D$,$\cal T$,$g$) is
Conformally Flat if, for each $m\in M$ there exists a neighbourhood
$\cal U$ of m, and a function $\Omega\in C^\infty(\cal U)$ such that
$\overline {\bf K} = 0$, where $\overline {\bf K}$ is the curvature associated
to $\overline{g} = e^{2\Omega}g$ (and to the connection given by the above
Theorem) }.

\medskip

\subsection{Theorem}

{\it If codim${\cal D} = n+1 \leq 2$, then $\cal T$ is always conformally
flat. For $n+1 \geq 4$, $\cal T$ is conformally flat iff ${\bf C} = 0$}

\medskip

{\bf Note}.

\medskip

{\small The case $n+1=3$ requires a special treatment (see
Th.4.3 in [10]).}

\medskip

Now we introduce some more definitions and computations, that we will use
later in the quantization program. First note that ${\bf K}_\parallel = 0$
implies that we can choose an orthonormal longitudinal parallel
transversal frame (in short, a OLPT frame) $E_a$, i.e.: $g(E_a,E_b)=0$,
for $a\not=b$ ; $g(E_a,E_a)=\epsilon_a (=\pm 1)$ and $\nabla_X E_a = 0,
\forall X\in \Gamma \cal D$.

In the following $E_a$ always designate such an OLPT frame for
$\Gamma\cal T$, and $\nabla$ the connection given by (\ref{54}). If $X_\alpha$
is a local frame for $\Gamma\cal D$, then (\ref{55}) gives:

$$0 ={\nabla}_{X_\alpha} E_a = (L_\alpha E_a)_\perp +1/2 \,
\lambda(X_\alpha)E_a$$
and, by (\ref{20}), with $\lambda = -\Theta$:

\begin{equation}
L_\alpha E_a = 1/2 \,  C_\alpha E_a + {\omega_\alpha^\beta} (E_a)X_\beta
\label{56}
\end{equation}

\subsection{Definition}

{\it We define the ``Ricci Tensor" }{\bf Ric }$\in {\cal D} ^{1,1}$ {\it by
the formula}:

\begin{equation}
{\bf Ric}(X;Q) = ({\bf c}{\bf K})(X;Q) = \sum_a \epsilon_a {\bf
K}(X,E_a;Q,E_a) \label{57}
\end{equation}

\medskip

\subsection{Lemma}

${\bf Ric}(X;Q) = n/2 \ \  d\Theta(X,Q)$.

\medskip

{\bf Proof}...

\medskip

{\small

We make use of the following identity from [10]:

$$cycl sum \, {\bf K}(X,Y;Z_\perp,.) = -1/2 \, cycl
sum \, d\Theta(X,Y)g(Z_\perp,.)$$
to compute ${\bf K}(X,E_a;Q,E_a)$. We have that:

$${\bf K}(X,E_a;Q,E_a) = -1/2 d\Theta(X,E_a)g(Q,E_a) - 1/2\epsilon_a
d\Theta(Q,X)$$

\medskip

Multiplying by $\epsilon_a$ and summing over $a=1,...,n$, we get easily
the result, using $\sum_a \epsilon_a g(Q,E_a)E_a = Q$ ////.}

\medskip

\subsection{Definition}

{\it For } $Q\in\Gamma\cal T$ {\it we define the ``Transversal Divergence
of} Q", {\it as the function}:

\begin{equation}
{\bf div_{\perp}}Q = \sum_a \epsilon_a g(\nabla_{E_a}Q,E_a)   \label{58}
\end{equation}

\medskip

\subsection{Lemma}

\begin{equation}
X_\alpha.{\bf div_{\perp}}U_\perp = C_\alpha {\bf div_{\perp}}U_\perp + (n-1)/2
\, d\Theta(X_\alpha,U_\perp)  \label{59}
\end{equation}

\medskip

{\bf Proof}...

\medskip

{\small

Omiting the symbols $\sum_a$ and  $\perp$ in $U_\perp$ and ${\bf div_{\perp}}$,
we have:

\begin{eqnarray*}
X_\alpha.{\bf div}{U}&=& X_\alpha.\epsilon_a
g(\nabla_{E_a}U,E_a)\\
&=& \epsilon_a
g(\nabla_{X_\alpha}\nabla_{E_a}U,E_a) +
\epsilon_ag(\nabla_{E_a}U,\nabla_{X_\alpha}E_a)\\
&=& {\epsilon_a}g(\nabla_{E_a}\nabla_{X_\alpha}U + \nabla_{[X_\alpha,E_a]}U
+ {\bf R}(X_\alpha,E_a)U,E_a)\\
&=& {\epsilon_a} g(\nabla_{E_a}(1/2{C_\alpha} U),E_a)
+ \epsilon_a g(\nabla_{1/2C_\alpha E_a+\omega_\alpha
^\beta(E_a)X_\beta}U,E_a) +\\
&&\quad {\bf Ric}(X_\alpha;U)
\end{eqnarray*}
where we have used (\ref{56}) for $L_\alpha E_a = [X_\alpha,E_a]$ and the fact
that $\nabla_{X_\alpha} U = 1/2 C_\alpha U$ (use (\ref{55}) together with
({24})). So, computing we obtain:

\begin{eqnarray*}
X_\alpha.{\bf div}{U}
&=& 1/2 {\epsilon_a} g((E_a.C_\alpha)U + C_\alpha\nabla_{E_a}U,E_a) +
1/2{\epsilon_a} g(C_\alpha \nabla_{E_a}U,E_a) +\\
&&\quad {\epsilon_a}{\omega_\alpha^\beta}(E_a) g(\nabla_{X_\beta}U,E_a) +
{\bf Ric}(X_\alpha,U)\\
&=&{1/2 \epsilon_a}d{C_\alpha}(E_a) g(U,E_a) +
{1/2 \epsilon_a} C_\alpha g(\nabla_{E_a}U,E_a) +\\
&&\quad   1/2{\epsilon_a} C_\alpha g(\nabla_{E_a}U,E_a) + {\epsilon_a}
{\omega_\alpha^\beta}(E_a) g(1/2 {C_\beta U},E_a) +\\
&&\quad {\bf Ric}(X_\alpha;U)\\ &=& 1/2 d{C_\alpha}(U) + 1/2 C_\beta
{\omega_\alpha^\beta}(U) + C_\alpha {\bf div}U + {\bf Ric}(X_\alpha;U)\\
&=& 1/2 {d\Theta}(U,X_\alpha) + C_\alpha {\bf div}U + {\bf
Ric}(X_\alpha;U)
\end{eqnarray*}
where we have used the following:
\begin{eqnarray}
U.C_\alpha&=& U(\Theta(X_\alpha)) \nonumber \\
&=& (L_U \Theta)(X_\alpha) +
\Theta([U,X_\alpha]) \nonumber \\
&=& (i_U d\Theta +d i_U \Theta)(X_\alpha)
+ \Theta(-C_\alpha U - C_\alpha^\beta X_\beta) \nonumber \\
&=& d\Theta(U,X_\alpha) - C_\beta C_\alpha^\beta  \label{60}
\end{eqnarray}
with $C_\alpha^\beta = \omega_\alpha^\beta(U)$.

 Finally, using Lemma 4.8, we obtain:
\begin{eqnarray*}
X_\alpha.{\bf div}U &=& C_\alpha {\bf div}U + 1/2
d\Theta(U,X_\alpha) + n/2 d\Theta(X_\alpha,U)\\
&=& C_\alpha {\bf div}U +(n-1)/2 d\Theta(X_\alpha,U). ////
\end{eqnarray*}}

\medskip

Finally, we want to define a kind of Scalar Curvature for the connection
$\nabla$, given by (\ref{54}).

\subsection{Definition}

{\it We define the ``Scalar Curvature of $\nabla$" by:}
\begin{eqnarray*}
{\bf S} &=& {\bf c}\circ{\bf c} \, K\\
&=&{\bf c}\,{\bf Ric}.
\end{eqnarray*}

\medskip

We will need the longitudinal derivative of ${\bf S}$, $X_\alpha.{\bf
S}$. For this, we first recall the Bianchi Identity for the connection
$\nabla$ on $({\cal T},g)$ (see [13], pag.89):

\begin{equation}
cycl_{(U,V,W)} \, {\nabla_U}{\bf R}(V,W) = cycl_{(U,V,W)} \, {\bf R}([U,V],W)
\label{61}
\end{equation}

\medskip

Recall that $({\nabla}_U{\bf R}(V,W))(Q) = {\nabla_U}({\bf R}(V,W)Q) - {\bf
R}(V,W)({\nabla_U}Q)$. Taking the inner product with another
$S\in\Gamma\cal T$, and using the properties of $\nabla$, we easily see
that we can write Bianchi's Identity in the form:
\begin{eqnarray}
&&cycl_{(U,V,W)}
(U.{\bf K}(V,W;Q,S)
-
{\bf K}(V,W;Q,{\nabla_U}S)
-
{\bf K}(V,W;{\nabla_U}Q,S))= \nonumber\\
&&\quad =
cycl_{(U,V,W)}
{\bf K}([U,V],W;Q,S))  \label{62}
\end{eqnarray}

\medskip

Now put $U=X_\alpha$, $V=E_c$, $W=E_d$, $Q=E_a$, $S=E_b$ in (\ref{62}) and
compute the contractions in the pairs of indices (d,b) and (c,a),
respectivelly, using formula (\ref{60}) and the fact that ${\bf N}_\perp = 0$
(see definition 4.2(ii)). After a tedious calculation we conclude that:

\subsection{Lemma}

\begin{eqnarray}
{X_\alpha}{\bf S} &=& {C_\alpha}{\bf S} - n(2{V_\alpha ^\beta}
.C_\beta + C_\beta {\bf div_{\perp}} {V_\alpha ^\beta} +\nonumber\\
&&\quad  C_\beta
{\omega_\gamma ^\beta}(V_\alpha ^\gamma) + {\bf \Delta} C_\alpha) \label{63}
\end{eqnarray}
{\it where}:

\begin{equation}
{V_\alpha ^\beta} = {\sum_a} {\epsilon_a} {\omega_\alpha
^\beta}(E_a) E_a = \eta^{ac} {\omega_\alpha ^\beta}(E_c) E_a  \label{64}
\end{equation}

\medskip
{\it is the transversal vector field metric-equivalent to the 1-form
$\omega_\alpha ^\beta$, and ${\bf \Delta} C_\alpha)$ is the
``{\it Transversal Laplacian}" of $C_\alpha$, defined by}:

\subsection{Definition}

{\it For a function} $\phi \in C^\infty(M)$ {\it we define its
``Transversal Laplacian" ${\bf \Delta}$ by}:

\begin{equation}
{\bf \Delta} \phi = {\bf div_{\perp}} \, {\bf grad}_\perp \phi  \label{65}
\end{equation}

\medskip
where ${\bf grad}_\perp \phi = \sum_a \epsilon_a d{\phi}(E_a)E_a = {\eta
^{ab}}(E_a.\phi)E_b$ is the ``{\it Transversal Gradient}" of $\phi$.

\medskip

 We can compute that:

\begin{eqnarray}
{\bf \Delta} \phi &=& \sum_a \epsilon_a g({\nabla_{E_a}}{\bf grad}_\perp
\phi,E_a) \label{66}\\
&=& \eta^{ab}(E_aE_b - \nabla_{E_a}E_b)\phi  \label{67}
\end{eqnarray}

\section{Quantization}

When we face the problem of quantizing a gauge system, two different
approaches are conceivable, in principle. We can reduce the gauge system
to the physical system (if possible) and quantize or, either, quantize the
gauge system directly and then reduce by some ``{\it Quantum Reduction
Process}". Hopefully, these two processes must be consistent, i.e.,
schematically
the following diagram must commute: (see discussion in [8] and [5])
\medskip
\begin{center}
\begin{picture}(200,100) (0,-10)
\put (0,0) {$\begin{array}{c}\hbox{Quantum}\\
\hbox{Gauge System}
\end{array}$}
\put(70,0){\vector (1,0){100}}
\put(100,6) {$\begin{array}{c}\hbox{Quantum}\\
\hbox{reduction}
\end{array}$}
\put(200,0){$\begin{array}{c}\hbox{Quantum}\\
\hbox{Physical System}
\end{array}$}
\put(0,80){Gauge System}
\put(70,80){\vector (1,0){100}}
\put(100,86)
{$\begin{array}{c}\hbox{Classical}\\
\hbox{reduction}
\end{array}$}
\put(200,80){Physical System}
\put(6,50){Quant.}
\put(200,50){Quant.}
\put(45,75){\vector (0,-1){65}}
\put(245,75){\vector (0,-1){65}}
\end{picture}
\end{center}
\medskip

As we have seen in the preceeding sections, the reduced physical system
(when exists) is characterized by a conformal class of physical
superhamiltonians:
$${h} = {e^{w}(g + J_u +v)}   \quad  w\in C^\infty(m)$$

Moreover, the dynamical constraint remains after reduction: $h=0$. So,
the proper way of quantizing this relativistic physical system is through
a ``{\it Conformal Klein-Gordon}" type equation.

\medskip

{\bf Note}.

\medskip

{\small  Recall that an equation of type ${\cal D}_g \Phi = 0$, for a field
$\Phi$ (where ${\cal D}_g$ is an operator constructed from the metric
$g$), is called ``{\it Conformally Invariant}" if there exists a number
$k\in \Re$ (called the ``{\it Conformal Weight of the Equation, or of the
Field}") such that $\Phi$ is a solution of ${\cal D}_g \Phi = 0$ if and
only if $\overline {\Phi} = e^{k\Omega} \Phi$ is a solution of ${\cal
D}_{\overline {g}}\overline{\Phi} = 0$, where  ${\cal
D}_{\overline {g}}$ is the same operator but now constructed from the
metric $\overline {g} = e^{\Omega} g$, (see [17]).}

\medskip

Now we define The ``{\it Quantum Operators}" corresponding respectivelly to
$g$, $J_u$ and $v$, by:

\begin{eqnarray}
\hat {\bf g} &=& - \Delta_c \nonumber \\
&=&  - \Delta_g - \xi{\bf S}_g   \label{68}\\
\hat {\bf J}_u  &=& 1/i \, (u + 1/2 \,{\bf div}_g u)  \label{69}\\
\hat {\bf v} &=& v   \label{70}
\end{eqnarray}
acting on $C^\infty(m)$, where $\Delta_c \equiv \Delta_g + \xi {\bf S}_g$ is
the ``{\it Conformal Laplacian}" of the metric $g$ (here, $\Delta_g$ and ${\bf
S}_g$ are, respectively, the usual Laplacian and Scalar Curvature of
$g$), and $\xi = (n-1)/4n$.

Then it's easy to see that the ``{\it Conformal Klein-Gordon Equation}":

\begin{equation}
\hat {\bf h} \psi = (\hat {\bf g} + \hat {\bf J}_u +\hat {\bf v})\psi = 0
\label{71}
\end{equation}
is conformally invariant with weight $k = (n-1)/4$.

Of course, now we must face the problem of the possibility of
construction of an Hilbert Space from solutions of (\ref{71}), as well as the
interpretation of the theory  (one-particle, second quantization, etc.)
(see [7] and [2], for which we defer the discussion of these
subjects).

What about ``{\it Constraint Quantization}" ? Here we adopt Kucha\v r's
philosophy, reinforced by the essential unique character of the objects
defined in section 4, namely, the connection $\nabla$, the corresponding
scalar curvature ${\bf S}$ (Definition 4.11) and transversal laplacian ${\bf
\Delta}$ (Definition 4.13).

So, we quantize the supermomentum constraints $H_\alpha$ by the following
operators:

\begin{equation}
\hat {\bf H}_\alpha = 1/i \, (X_\alpha - kC_\alpha) \label{72}
\end{equation}
and the transversal superhamiltonian $H_\perp$, by:

\begin{equation}
\hat {\bf H}_\perp = \hat {\bf g} + \hat {\bf J}_{U_\perp} + \hat {\bf V}
\label{73}
\end{equation}
with
\begin{eqnarray}
\hat {\bf g} &=& - {\bf \Delta} - \xi {\bf S}  \label{74}\\
\hat {\bf J}_{U_\perp} &=& 1/i \, (U_\perp + 1/2 \, {\bf
div_{\perp}}{U_\perp})   \label{75}\\
\hat {\bf V} &=& V    \label{76}
\end{eqnarray}
acting on $C^\infty(M)$.

\medskip

Then the classical constraints are imposed at quantum level as the ``{\it
Quantum Constraints}":

\begin{equation}
\hat {\bf H}_\alpha \psi = 0 = \hat
{\bf H}_\perp \psi  \ \ \ \ \ \ \ \   \psi \in
C^\infty(M)
\label{77}
\end{equation}

As we have said, this ``{\it Quantum Reduction Process}" must be
consistent with the first approach to quantization (see discussion in [8]
and [5]). This is implied by the following Lemmas and Theorems.

\subsection{Theorem}

$$1/i \, [\hat {\bf H}_\alpha,\hat {\bf H}_\beta] = \Gamma_{\alpha\beta}^\gamma
\hat {\bf H}_\gamma$$
{\it where} $\Gamma_{\alpha\beta}^\gamma \in C^\infty(M)$ {\it are given by
(\ref{9})}.

\medskip

{\bf Proof}

\medskip

{\small

Compute, using definitions and (\ref{32}). ////}

\medskip

\subsection{Lemma}

$$1/i \,[\hat {\bf J}_{U_\perp},\hat {\bf H}_\alpha] = C_\alpha \hat
{\bf J}_{U_\perp} + \omega_\alpha^\beta(U_\perp) \hat {\bf H}_\beta$$
{\it where} $C_\alpha = \Theta(X_\alpha)$ {\it and}
$\omega_\alpha^\beta(U_\perp)$ is given by (\ref{24}).

\medskip

{\bf Proof}

\medskip

{\small

Using Definition (\ref{75}) we compute the commutator in the LHS of the above
equation, and conclude that:
\begin{eqnarray*}
1/i \,[\hat {\bf J}_{U_\perp},\hat {\bf H}_\alpha] \phi &=& (C_\alpha \hat
{\bf J}_{U_\perp} + \omega_\alpha^\beta(U_\perp) \hat {\bf H}_\beta)\psi +
k/i\,(U_\perp.C_\alpha + \omega_\alpha^\beta(U_\perp) C_\beta) + \\
&&\quad 1/2\,(X_\alpha.{\bf div_{\perp}}{U_\perp} - C_\alpha
{\bf div_{\perp}}{U_\perp}))\psi
\end{eqnarray*}

However, by (\ref{60}), $U_\perp.C_\alpha + C_\beta
\omega_\alpha^\beta(U_\perp) = d\Theta(U_\perp,X_\alpha)$, while Lemma 4.10
gives $X_\alpha.{\bf div_{\perp}}{U_\perp} = C_\alpha {\bf
div_{\perp}}{U_\perp} +(n-1)/2\, d\Theta(X_\alpha,U_\perp)$, and so we see
that the last sum in the RHS is zero.////}

\subsection{Lemma}

$$1/i \, [\hat{\bf g},\hat {\bf H}_\alpha] = C_\alpha \hat{\bf g} +
2(\hat{\bf V}_\alpha^\beta - i/2 \, \omega_\gamma^\beta (V_\alpha^\gamma))\hat
{\bf H}_\beta$$
{\it where} $V_\alpha^\beta$ {\it is defined in (\ref{64}), and, as usual (see
[1])}:
\begin{equation}
\hat{\bf V}_\alpha^\beta = 1/i\,(V_\alpha^\beta +1/2 \,
{\bf div_{\perp}}{V_\alpha^\beta}) \label{78}
\end{equation}
{\it with} ${\bf div_{\perp}}{V_\alpha^\beta}$ {\it as in Definition 4.9,
equation (\ref{58})}.

\medskip

{\bf Proof}

\medskip

{\small

First we compute that:

\begin{equation}
1/i\, [\hat{\bf g},\hat{\bf H}_\alpha] = [{\bf \Delta},X_\alpha]
-k[{\bf \Delta},C_\alpha] + \xi[{\bf S},X_\alpha]  \label{79}
\end{equation}
and it's easy to see that:
\begin{equation}
[{\bf \Delta},C_\alpha] = 2i \, \widehat {\bf grad}_\perp C_\alpha = 2({\bf
grad}_\perp C_\alpha + 1/2\, {\bf \Delta} C_\alpha)  \label{80}
\end{equation}
and
\begin{equation}
[{\bf S},X_\alpha] = - X_\alpha.{\bf S}  \label{81}
\end{equation}
So we must compute $[{\bf \Delta},X_\alpha]$. For this, we use formulas
(\ref{66}-\ref{67}) for ${\bf \Delta}\phi$ and compute that:
\begin{equation}
[{\bf \Delta},X_\alpha]\phi = {\eta}^{ab}(E_aE_bX_\alpha - X_\alpha E_aE_b)\phi
+
{\eta}^{ab}(X_\alpha{\nabla_{E_a}}E_b - {\nabla_{E_a}}E_bX_\alpha)\phi
\label{82}
\end{equation}

Now we commute the derivatives and uses systematically (\ref{56}), to get:

\begin{eqnarray*}
 X_\alpha E_a E_b \phi &=& E_a E_bX_\alpha \phi + C_\alpha E_a E_b \phi +
1/2(E_a.C_\alpha) E_b \phi +\\
&&\quad \omega_\alpha^\beta (E_b) E_a X_\beta \phi +
(E_a.\omega_\alpha^\beta(E_b))X_\beta \phi + \omega_\alpha^\beta
(E_a) E_b X_\beta \phi +\\
&& \quad 1/2C_\beta \omega_\alpha^\beta (E_\alpha) E_b \phi +
\omega_\alpha^\beta (E_a)\omega_\beta^\gamma (E_b)X_\gamma \phi
\end{eqnarray*}
and:
\begin{eqnarray*}
(X_\alpha{\nabla_{E_a}}E_b - {\nabla_{E_a}}E_b X_\alpha) \phi &=&
(C_\alpha {\nabla_{E_a}}E_b + {\bf R}(X_\alpha,E_a)E_b +\\
&&\quad \omega_\alpha^\beta({\nabla_{E_a}}E_b)X_\beta)\phi
\end{eqnarray*}

Substituting these last two identities in (\ref{82}), we obtain:
\begin{eqnarray*}
[{\bf \Delta},X_\alpha]\phi &=& -C_\alpha {\bf \Delta} \phi - (2 V_\alpha^\beta
+
{\bf div_{\perp}}V_\alpha^\beta + \omega_\gamma^\beta(V_\alpha^\gamma))X_\beta
\phi -\\ &&\quad 1/2({\bf grad}_\perp C_\alpha + C_\beta V_\alpha^\beta - 2
\eta^{ab}{\bf R}(X_\alpha,E_a)E_b)
\end{eqnarray*}
Now we use the fact:
\begin{eqnarray*}
\eta^{ab}{\bf R}(X_\alpha,E_a)E_b &=& \eta^{ab} {\sum_c} \epsilon_c
{\bf K}(X_\alpha,E_a;E_b,E_c)E_c \\ &=& -{\sum_c}\epsilon_c \eta^{ab}
{\bf K}(X_\alpha,E_a;E_b,E_c)E_c \\  &=& n/2 {\sum_c}\epsilon_c
(E_c.C_\alpha + C_\beta \omega_\alpha^\beta(E_c))E_c \\
 &=& n/2\, ({\bf grad}_\perp C_\alpha + C_\beta V_\alpha^\beta)
\end{eqnarray*}
where we have used (\ref{60}), Lemma 4.10  and (\ref{64}).

By [13] (pag.151), we have:
\begin{eqnarray*}
{\bf div_{\perp}} {V_\alpha^\beta} &=& {\bf div_{\perp}}
{\omega_\alpha^\beta}\\
&=& \eta^{ab}(\nabla_{E_a}{\omega_\alpha^\beta})(E_b) \\
&=& \eta^{ab}(E_a.{\omega_\alpha^\beta}(E_b) -
{\omega_\alpha^\beta}(\nabla_{E_a} E_b))
\end{eqnarray*}
and so, substituting and calculating, we have:
\begin{eqnarray*}
[{\bf \Delta},X_\alpha]\phi &=& - C_\alpha {\bf \Delta} \phi - C_\alpha \xi
{\bf S}\phi +2(\hat{\bf V}_\alpha^\beta - i/2 \, {\omega_\gamma^\beta}
(V_\alpha^\gamma)) \hat H_\beta \phi - \\
&&\quad k(2{V_\alpha^\beta}.C_\beta + C_\beta {\bf div_{\perp}} V_\alpha^\beta
+
C_\beta {\omega_\gamma^\beta}(V_\alpha^\gamma)) \phi + \\
&&\quad 2k{{\bf grad}_\perp} C_\alpha \phi + C_\alpha \xi {\bf S} \phi
\end{eqnarray*}

By (\ref{80}) - (\ref{82}), we have:
\begin{eqnarray*}
1/i [\hat{\bf g},\hat {\bf H}_\alpha] \phi &=& C_\alpha \hat{\bf g} \phi +
2(\hat{\bf V}_\alpha^\beta - i/2\,
\omega_\gamma^\beta(V_\alpha^\gamma))\hat {\bf H}_\beta \phi \\
&=& - k(2 V_\alpha^\beta.C_\beta + C_\beta {\bf div_{\perp}} V_\alpha^\beta +
C_\beta \omega_\gamma^\beta(V_\alpha^\gamma))\phi +\\
&&\quad C_\alpha \xi {\bf S} \phi - k {\bf \Delta} C_\alpha - \xi(X_\alpha.{\bf
S})\phi
\end{eqnarray*}
and so, it suffices to prove that:
\begin{eqnarray*}
\xi X_\alpha.{\bf S} &=& \xi C_\alpha {\bf S} - k(2 V_\alpha^\beta.C_\beta +
C_\beta {\bf div_{\perp}} V_\alpha^\beta +\\
&&\quad C_\beta \omega_\gamma^\beta(V_\alpha^\gamma) + \Delta C_\alpha)
\end{eqnarray*}
which is preciselly Lemma 4.12. ////}

\medskip

Now, collecting the above two Lemmas, together with (13), we finally have
the following:

\subsection{Theorem}

$$1/i\,[\hat {\bf H}_\perp,\hat {\bf H}_\alpha] = C_\alpha \hat {\bf H}_\perp +
{\bf C_\alpha^\beta} \hat {\bf H}_\beta$$
{\it where}:

$${\bf C_\alpha^\beta} = 2 \hat{\bf V}_\alpha^\beta - i
\omega_\gamma^\beta({\bf V}_\alpha^\gamma) + \omega_\alpha^\beta(U_\perp)$$

\bigskip

So, Theorems 5.1 and 5.4 show that we have a closed commutator algebra of
quantum operator constraints, with the structure functions appearing on the
left
of that quantum operators. This implies the consistency of the above
quantization process. Moreover we recover the physical theory described
by the ``{\it Conformal Klein-Gordon equation}" (\ref{71}). In fact, start with
a wave function $\Psi \in C^\infty (M)$ that solves the quantum
constraints (\ref{77}). Assume also that the foliation $\cal F$ is simple, and
that $\Theta$ is $d_{\cal F}-exact$, i.e., $\exists \Omega \in C^\infty
(M)$ such that $\Theta=-d_{\cal F} \Omega$. Then we easily prove that
the rescaled wave function $\overline {\Psi} = e^{k{\Omega}}\Psi$ is a
basic function and so, descends to a wave function $\overline {\psi}
\in C^\infty (m)$. Now, as in [6], we can prove that the transversal
superhamiltonian $\hat {\bf H}_\perp$ has conformal weight 1, when acting on
wave functions of conformal weight $k$, i.e.:
\begin{equation}
\hat {\overline {\bf H}}_\perp \overline {\Psi} = e^{(k+1)\Omega}\hat {\bf
H}_\perp \Psi \label{83}
\end{equation}
where $\overline {\Psi} = e^{k{\Omega}}\Psi$, and $\hat {\overline{\bf
H}}_\perp$ is the transversal superhamiltonian constructed from the rescaled
metric $\overline {g} = e^{\Omega} g$. So we see that (\ref{77}) $\Rightarrow
\hat {\bf H}_\perp \Psi = 0 \Rightarrow \hat {\overline {\bf H}}_\perp
\overline \Psi = 0$, and, since $\overline \Psi$ is basic, this equation
implies that:
\begin{equation}
\hat {\overline {\bf h}} \, \overline {\psi} = 0
\end{equation}
where $\overline{\psi}$ is the induced wave function on $m$. In this way,
we recover the physical conformal Klein-Gordon equation (\ref{71}).

\medskip

\medskip

\medskip
\medskip

\section{Appendix}

{\small

Hereafter we adopt the following notations, related to decompositions
(\ref{15})-(\ref{18}):

$Q^A$ are local coordinates on $M$

$\partial_A = {\partial}$/${\partial Q}^A$, and  $\partial^A = dQ^A$,
$A=0,...,N$

$X_\alpha = X_\alpha^A\partial_A$ , a local frame field for $\Gamma{\cal
D}$, $\alpha = 1,...,v$

$Q_a = Q_a^A\partial_A$ a local frame field for $\Gamma{\cal T}$, $a =
0,...,n$

$(\theta^\alpha,\theta^a)$, a dual coframe for ${\cal D}\oplus{\cal
T}$, so that $\theta^\alpha \in \Gamma{\cal D}^*$, $\theta^a \in
\Gamma{\cal T}^* \cong \Gamma{\cal D}^o$ (we may assume that $\theta^a$
are closed 1-forms), $\theta^\alpha(X_\beta)=\delta_\beta^\alpha$,
$\theta^a(Q_b)={\delta_b^a}$, $\theta^\alpha(Q_a)=0=\theta^a(X_\alpha)$.

$G = G^{AB}\partial_A\otimes\partial_B$ , a contravariant metric on
$M$.

$\tilde g=G|({\cal D}^o \cong {\cal T}^*)$

The projector $P_\perp: TM \to {\cal T}$ can be writen in the form:

\begin{eqnarray*}
P_\perp &=& Q_a \otimes \theta^a(.)  \\
&=& Q_a^B Q_A^a \partial_B  \otimes \partial^A \\
&=& {\cal P}_A^B \partial_B \otimes \partial^A
\end {eqnarray*}
and the transversal metric $G_\perp$ is:

\begin{eqnarray*}
G_\perp &=& P_\perp \otimes P_\perp (G) \\
&=& Q_a \otimes \theta^a \otimes Q_b \otimes \theta^b (G^{AB}
\partial_A \otimes \partial_B)\\
&=& G^{AB} Q_a^C Q_A^a Q_b^E Q_B^b \partial_C \otimes \partial_E \\
&=& G^{AB} {\cal P}_A^C {\cal P}_B^E \partial_C \otimes
\partial_E.
\end{eqnarray*}

{}From
\begin{eqnarray*}
{\tilde g} ^{ab} &=& G(\theta^a,\theta^b) \\
&=& G^{AB} \partial_A \otimes \partial_B (Q_C^a \partial^C,Q_D^b
\partial^D) \\
&=& G^{AB} Q_A^a Q_B^b,
\end{eqnarray*}
we can also write:
\begin{equation}
G_\perp = {\tilde g}^{ab} Q_a^C Q_b^E \partial_C \otimes \partial_E \label{85}
\end{equation}

Now it's easy to see that:

\begin{equation}
L_\alpha \theta^a \in {\cal D}^o \cong {\cal T}^*  \quad   \forall \alpha, a
\label{86}
\end{equation}

We compute now $L_\alpha P_\perp = L_\alpha (Q_a \otimes \theta^a)$: first we
decompose $ L_\alpha Q_a = t_{\alpha a}^b Q_b + l_{\alpha a} ^\beta X_\beta$.
However, we have:
$$0 = L_\alpha <\theta^b,Q_a> = <L_\alpha \theta^b,Q_a> +
<\theta^b,L_\alpha Q_a>$$
 and by (\ref{85}), we can put $L_\alpha \theta^b = r_{\alpha d}^b \theta^d$,
and deduce, from the last equation, that $t_{\alpha a}^b = - r_{\alpha a}^b$.
So, reuning these information, we easily compute that:

\begin{equation}
L_\alpha P_\perp = l_{\alpha a}^\beta X_\beta \otimes \theta^a \label{87}
\end{equation}
where $l_{\alpha a}^\beta = \theta^\beta(L_\alpha Q_a)$.

\medskip

Now we compute $L_\alpha U_\perp$, with $U_\perp = P_\perp (U)$ and $U \in
{\cal X}(M)$. By (\ref{22}):
\begin{eqnarray*}
L_\alpha U_\perp &=& L_\alpha (P_\perp \bullet U) \\
&=& (L_\alpha {P_\perp}) \bullet U + {P_\perp} \bullet (L_\alpha U) \\
&=& (l_{\alpha a}^\beta X_\beta \otimes \theta^a)(u^a Q_a + u^\beta
X_\beta) + \\
&&\quad P_\perp(C_\alpha + long. vector)  \ \ \ \ \  {\hbox{by (\ref{12})}} \\
&=& l_{\alpha a}^\beta u^a X_\beta + C_\alpha U_\perp,
\end{eqnarray*}
and recalling the definition of the 1-form $\omega_\alpha^\beta$ in
(\ref{23}), it's easy to see that: $\omega_\alpha^\beta = l_{\alpha a}^\beta
\theta^a$, and so we can write $L_\alpha U_\perp$ also in the form (\ref{24}).

\medskip

Now we compute $L_\alpha G_\perp$. Again by (\ref{22}), we have, with ${\cal
P}_\perp = P_\perp \otimes P_\perp$:
\begin{eqnarray*}
L_\alpha G_\perp &=& L_\alpha ({\cal P}_\perp \bullet G) \\
&=& {\cal P}_\perp \bullet (L_\alpha G) + (L_\alpha {\cal P}_\perp)
\bullet G
\\ &=&  {\cal P}_\perp \bullet (C_\alpha G + terms \, in \, {\bf I}) + L_\alpha
(P_\perp \otimes P_\perp) \bullet G \\
&=& C_\alpha G_\perp + (L_\alpha P_\perp \otimes P_\perp + P_\perp
\otimes L_\alpha P_\perp) \bullet G \\
&=& C_\alpha G_\perp + l_{\alpha a}^\beta X_\beta \otimes \theta^a
\otimes Q_b \otimes \theta^b + Q_b \otimes \theta^b \otimes \\
&&\quad \otimes l_{\alpha a}^\beta X_\beta \otimes \theta^a ({\tilde g}^{cd}
Q_c \otimes Q_d)\\
&=& C_\alpha G_\perp + l_{\alpha a}^\beta {\tilde g}^{ab} X_\beta \otimes
Q_b + l_{\alpha a}^\beta {\tilde g}^{ba} Q_b \otimes X_\beta
\end{eqnarray*}
where we have used (\ref{11}) and the fact that $\theta^a \in {\cal D}^o$. Now,
let $V_\alpha^\beta = G(\omega_\alpha^\beta,.) \in \Gamma {\cal T}$ the
transversal vector field, metric-equivalent to the transversal 1-form
$\omega_\alpha^\beta$, given by (\ref{23}). We compute that $ V_\alpha^\beta =
l_{\alpha a}^\beta {\tilde g}^{ab} Q_b$, and so we see that we can write
$L_\alpha G_\perp$ in the form (\ref{25}).

As in section 3, we assume that ${\tilde g} = G|{\cal D}^o = G_\perp |{\cal
D}^o$ is nondegenerate, and let $g$ be the corresponding transversal covariant
metric on $\cal T$. We want to compute $L_\alpha g$. For this, we first note
that $L_\alpha g$ must be a covariant transversal vector field, since $g =
g_{ab} \theta^a \otimes \theta^b$ and $L_\alpha \theta^a \in {\cal D}^o$.
So, we compute $(L_\alpha {\tilde g})|{\cal T}$, since it sufices to compute
$L_\alpha g$. We have:
\begin{eqnarray*}
(L_\alpha {\tilde g})(\theta^a,\theta^b) &=& (L_\alpha
G_\perp)(\theta^a,\theta^b) \\
&=& (C_\alpha G_\perp + {V_\alpha^\beta} \vee {X_\beta})(\theta^a,\theta^b)
\\ &=& C_\alpha G_\perp (\theta^a,\theta^b) \ \ \ \ \ {\hbox{since \,
$\theta^a \in {\cal D}^o$}} \\
&=& C_\alpha {\tilde g} (\theta^a,\theta^b) \ \ \ \ \ \ \ {\hbox{by
(\ref{85})}} \end{eqnarray*}
and so, defining $L_{\alpha}g$ as:
\begin{equation}
(L_{\alpha}g)(Q_a,Q_b)=g_{ac}g_{bd}(L_{\alpha}{\tilde g})(\theta ^a,\theta ^b)
\label{88}
\end{equation}
we conclude that $L_{\alpha}g=-C_{\alpha}g$. Notice that if $G$ is a covariant
metric on $M$, such that $G({\cal D},{\cal T})=0$, then the definition
(\ref{88}) of $L_{\alpha}g$ is equivalent to:
\begin{equation}
(L_{\alpha}g)(Q_a,Q_b)=(L_{\alpha}G)(Q_a,Q_b)  \label{89}
\end{equation}}

\newpage

{\bf Acknowledgments}

\medskip

I am extremelly grateful to J.Mour\~ao for many helpful comments and criticisms
which substantially improved the presentation.

\medskip
\medskip

{\bf References}

\medskip

[1]. Abraham R., Marsden J.E. ``{\it Foundations of Mechanics}" (2.nd
edition), Benjamin Cumming P. Company, 1978.

\medskip

[2]. Ashtekar A., Magnon A.. Quantum Fields in Curved
Spacetimes, {\it Proc.R.Soc.London A} 346 (1975) 375-394.

\medskip

[3]. Gotay M.J.. Constraints, reduction and quantization, {\it J. Math. Phys.}
27(8) (1986) 2051-2066.

\medskip

[4]. Guillemin V., Sternberg S. ``{\it Sympletic Techniques
in Physics}", Cambridge Un.Press, 1984.

\medskip

[5]. H\'ajic\~ek P., Kucha\v r K.. Constraint Quantization of
Parametrized Relativistic Gauge Systems in Curved Spacetimes, {\it Physical
Review D} 41, n.4 (1990) 1091-1104.

\medskip

[6]. H\'ajic\~ek P., Kucha\v r K..
Transversal Affine Connection and Quantization of Gauge Systems, {\it
J.Math.Phys.} 31(7) (1990) 1723-1732.

\medskip

[7]. Kucha\v r K.. Canonical Methods of Quantization, in
{\it ``Quantum Gravity"} (vol.2) Isham C.J., Penrose R. (eds), Oxford, 1980.

\medskip

[8]. Kucha\v r K.. Covariant Factor Ordering of Gauge Systems,
{\it Physical Review D} 34 (10) (1986) 3044-3057.

\medskip

[9]. Molino P. ``{\it Riemannian Foliations}", Progress in
Mathematics vol.73, Birkhauser, 1988.

\medskip

[10]. Montesinos A.. Conformal Curvature for the Normal Bundle
of a Conformal Foliation, {\it Ann.Inst.Fourier, Grenoble}, 32 (3) (1982)
261-274.

\medskip

[11]. Montesinos A.. On Certain Classes of Almost Product
Structures, {\it Michigan Math.J.} 30 (1983) 31-36.

\medskip

[12]. O'Neill B. ``{\it Semi-Riemannian Geometry}", Academic Press,
1983.

\medskip

[13]. Poor W.A. ``{\it Differential Geometric Structures}",
McGraw-Hill Book Company, 1981.

\medskip

[14]. Vaisman I. ``{\it Conformal Foliations}", Kodai Math. J. 2
(1979) 26-37.

\medskip

[15]. Vaisman I..Vari\'et\'es Riemannienes Feuillet\'ees
{\it Czech.Math.J.} 21 (96) (1971) 46-75.

\medskip

[16]. Vaisman I. ``{\it Cohomology and Differential Forms}",
Marcel Dekker, Inc., New York, 1973.

\medskip

[17]. Wald R.M. ``{\it General Relativity}", The University of Chicago
Press, 1984.

\end{document}